\documentclass[pre,aps,epsfig,amsmath,amssymb,preprint]{revtex4}
\usepackage{times}
\usepackage{epsf}
\usepackage{graphicx}
\usepackage{dcolumn}
\usepackage{bm}

\begin{document}
\draft

\date {\today}

\title{Structure of polydisperse inverse ferrofluids: \\ Theory and computer simulation}
\author{Y. C. Jian, Y. Gao and J. P. Huang\footnote{Corresponding author. Electronic address: jphuang@fudan.edu.cn}}
\address{Surface Physics Laboratory  and Department of Physics, Fudan University, Shanghai
200433, China}
\author{R. Tao}
\address{Department of Physics, Temple University, Philadelphia,
 Pennsylvania 19122}

\begin{abstract}

By using theoretical analysis and molecular dynamics simulations,
we investigate the structure of colloidal crystals formed by
nonmagnetic microparticles (or {\it magnetic holes}) suspended in
ferrofluids (called {\it inverse ferrofluids}), by taking into
account the effect of polydispersity in size of the nonmagnetic
microparticles. Such polydispersity  often exists in real
situations. We obtain an analytical expression for the interaction
energy of monodisperse, bidisperse, and polydisperse inverse
ferrofluids. Body-centered tetragonal ($bct$) lattices are shown
to possess the lowest energy when compared with other sorts of
lattices, and thus serve as the ground state of the systems. Also,
the effect of microparticle size distributions (namely,
polydispersity in size) plays an important role in the formation
of various kinds of structural configurations. Thus, it seems
possible to fabricate colloidal crystals by choosing appropriate
polydispersity in size.
\end{abstract}


\maketitle
\section{Introduction}

In recent years, inverse ferrofluids with  nonmagnetic colloidal
microparticles suspended in a host ferrofluid (also called
magnetic fluid)$^{1-3}$ have drawn considerable attention for its
potential application in its industrial applications and potential
use in biomedicine.$^{4-11}$ The size of the nonmagnetic
microparticles are about $1\sim 100\,$$\mu$m, which can be easily
made in experiments, such as polystyrene microparticles. The
inverse ferrofluid system can be modelled in a dipolar interaction
approximation. Here, the dipolar interaction approximation is
actually  the first-order approximation of multipolar interaction.
Because the nonmagnetic microparticles are much larger than the
ferromagnetic nanoparticles in a host ferrofluid, the host can
theoretically be treated as a uniform continuum background in
which the much larger nonmagnetic microparticles are embedded. If
an external magnetic field is applied to the inverse ferrofluid,
the nonmagnetic microparticles suspended in the host ferrofluid
can be seen to posses an effective magnetic moment but opposite in
direction to the magnetization of the host ferrofluid. As the
external magnetic field increases, the nonmagnetic microparticles
aggregate and form chains parallel with the applied magnetic
field. These chains finally aggregate to a column-like structure,
completing a phase transition process, which is similar to the
cases of electrorheological fluids and magnetorheological fluids
under external electric or magnetic fields. The columns can behave
as different structures like body-centered tetragonal ($bct$)
lattices, face centered cubic ($fcc$) lattices, hexagonal close
packed ($hcp$) lattices, and so on. In this work, we assume that
the external magnetic field is large enough to form different
lattice structures. The actual value of the external magnetic
field needed to form such structures is related to the volume
fraction of the nonmagnetic microparticles and the magnetic
properties of the host ferrofluid.

In this work, we shall use the dipole-multipole interaction
model$^{12}$ to investigate the structure of inverse ferrofluids. In
ref 12, Zhang and Widom discussed how the geometry of elongated
microparticles will affect the interaction between two droplets, and
introduced higher multipole moments' contribution by using a dipole
and multipole (dipole-multipole) interaction model to give a more
exact expression of interaction energy than using the dipole and
dipole (dipole-dipole) interaction model. The leading dipole-dipole
force does not reflect the geometry relation between the
microparticles nearby, while the dipole-multipole model includes the
contributions from the size mismatch  and is simpler and practical
than the multipolar expansion theory$^{13,14}$ in dealing with the
complex interaction between microparticles for its accuracy. Size
distributions can be regarded as a crucial factor which causes
depletion forces in colloidal droplets.$^{15}$ Even though
researchers have tried their efforts to fabricate monodisperse
systems for obtaining optimal physical or chemical
properties,$^{16,17}$ polydispersity in size of microparticles often
exists in real situations,$^{18-22}$ since the microparticles always
possess a Gaussian or log-normal distribution. Here we consider size
distributions as an extra factor affecting the interaction energy.
Polydisperse ferrofluid models are usually treated in a global
perspective using chemical potential or free energy
methods,$^{6,23,24}$ while the current model concerns the local
nature in the crystal background. A brief modelling is carried out
for the size distribution picture in the formation of crystal
lattices. The purpose of this paper is to use this model to treat
the structure formation in monodisperse, bidisperse, and
polydisperse inverse ferrofluids, thus yielding theoretical
predictions for the ground state for the systems with or without
microparticle size distributions (or polydispersity in size). It is
found that when the size mismatch is considered between the
microparticles, the interaction between them becomes complex and
sensitive to the different configurations. This method can also be
extended to other ordered configurations in polydisperse crystal
systems.

This paper is organized as follows. In Section~\ref{secII}, based
on the dipole-multipole interaction model, we present the basic
two-microparticle interaction model to derive interaction
potentials. In Sections~\ref{secIII}~and~~\ref{secIV}, we apply
the model to three typical structures of colloid crystals formed
in inverse ferrofluids, and then  investigate the ground state in
two different configurations by taking into account the effect of
size distributions. As an illustration, in Section~\ref{secV} we
perform molecular dynamics simulations to give a picture of the
microparticle size distribution in the formation of a $bct$
lattice in bidisperse inverse ferrofluids. The paper ends with a
discussion and conclusion in Section~\ref{secVI}.

\section{interaction model for two nonmagnetic microparticles}\label{secII}

We start by considering a simple situation in which two
nonmagnetic spherical microparticles (or called {\it magnetic
holes}) are put nearby inside a ferrofluid which is homogeneous at
the scale of a sphere in an applied uniform magnetic field $H$,
see Figure 1. The nonmagnetic microparticles create holes in the
ferrofluid, and corresponding to the amount and susceptibility of
the ferrofluid, they possess the effective magnetic moment, which
can be described by$^{25}$
$\textbf{m}=-\frac{\chi_f}{1+\frac{2}{3}\chi_f}V\textbf{H}=-\chi
V\textbf{H},$ where $\chi_f$ (or $\chi$) means the magnetic
susceptibility of the host (or inverse) ferrofluid. When the two
nonmagnetic microparticles placed together with distance $r_{ij}$
away, we can view the magnetization in one sphere (labelled as A)
is induced by the second (B). The central point of
dipole-multipole technique is to treat B as the dipole moment m at
the first place, and then examine the surface charge density
$\Sigma$ induced on the sphere A. From $\Sigma$ we can use the
multipole expansion (detailed discussion can be found in ref 26)
to obtain the multipole moment. When exchanging the status of A
and B, treating A as the dipole moment, the averaged force between
the two microparticles is thus obtained. For the perturbation of
the magnetic field due to the two microparticles, magnetization
$M$ in the microparticles become nonuniform, and they will obtain
multipole moments from mutual induction. However, the bulk
magnetic charge density still satisfies $\rho=\nabla\cdot
\textbf{M}=0$. So we need only study the surface charge
$\Sigma_n$,
\begin{equation}
\Sigma_n=\hat{{\bf n}}\cdot\textbf{M},
\end{equation}
where $\hat{{\bf n}}$ is the unit normal vector pointing outwards.
The magnetic multipole moments by surface charge density in
spherical coordinates can be written as,
\begin{equation}
q_{ln}=\int Y_{ln}^{*}(\phi,\varphi)r_1^l\Sigma_n{\rm d}S,
\end{equation}
where $r_1$ denotes the radius of the microparticle. All $n\neq0$
moments vanish due to rotational symmetry about the direction of
magnetization, so eq 2 can be rewritten as
\begin{equation}
A_{l}\equiv
q_{l0}=\sqrt{(2l+1)\pi}\int_{-1}^{1}P_{l}(\cos\phi)r_{1}^{l+2}\Sigma_n(\cos\phi){\rm
d}\cos\phi .
\end{equation}
We can expand the surface charge density in Legendre polynomials
in the spherical coordinates ($r$,$\phi$,$\varphi$),$^{12}$
\begin{equation}
\Sigma_n(\cos
\phi)=\frac{\chi}{r_{ij}^{3}}\sum^{\infty}_{l=1}(-1)^{l+1}l(l+1)\left(\frac{r_1}{r_{ij}}\right)^{l-1}P_l(\cos
\phi). \label{eq1}
\end{equation}
The $l\ge 2$ parts in eq~\ref{eq1} correspond to the effects of
multipole (that are beyond the dipole). Here we set spherical
harmonics $\int_{0}^{2\pi}{\rm d}\varphi\int_{0}^{\pi}\sin\phi
Y_{l^{'},n^{'}}^{\ast}(\phi,\varphi)Y_{l}(\phi,\varphi){\rm
d}\phi=\delta_{l^{'},l}\delta_{n^{'},n} $. The force between the
dipole moment $m$ and induced multipole moment $A_{l}$ can be
derived as
\begin{equation}
F_{D-M}=(-1)^l(l+1)(l+2)(\frac{4\pi}{2l+1})^{\frac{1}{2}}\frac{mA_{l}}{r_{ij}^{l+3}}\cos\phi
.
\end{equation}
In view of the orthogonal relation
$\int_{-1}^1P_l^nP_{l^{'}}^n{\rm
d}x=\frac{2}{2l+1}\frac{(l+n)!}{(l-n)!}\delta_{ll^{'}}$, we obtain
the interaction energy for the dipole-multipole moment
\begin{equation}
U_{D-M}=\frac{\pi}{4}\mu_fm_1m_2\sum_l\frac{\pi}{2}\chi\frac{l(l+1)^2}{2l+1}(r_1^{2l+1}+
r_2^{2l+1})\frac{1-3\cos^2\theta}{r_{ij}^{2l+4}}=\mu
\sum_lf(l)\frac{1-3\cos^2\theta}{r_{ij}^{2l+4}},
\end{equation}
where $\mu=\frac{\pi}{4}\mu_fm_1m_2$,
$f(l)=\frac{\pi}{2}\chi\frac{l(l+1)^2}{2l+1}(r_1^{2l+1}+
r_2^{2l+1})$, the suffix $D$ $(M)$ of force $F$ or energy $U$
stands for the dipole moment (multipole moment), and magnetic
permeability $ \mu_f=\mu_0(1+\chi_f)$ with
$\mu_0=4\pi\times10^{-7}$ H$\cdot$m$^{-1}$. Here $m_1$ and $m_2$
denote the effective magnetic moments of the two nonmagnetic
microparticles, which is induced by external field as dipolar
perturbation. $0\leq\theta\leq\frac{\pi}{2}$ is the angle between
the their joint line with the direction of external field, and
$\phi$ and $\varphi$ are both the spherical coordinates
($r$,$\phi$,$\varphi$) for one single nonmagnetic microparticle.
For typical ferrofluids, there are magnetic susceptibilities,
$\chi_f=1.9$ and $\chi=0.836$.$^{23}$ Because we consider the
$bct$, $fcc$ and $hcp$ lattices, the crystal rotational symmetry
in the $xy$ plane is fourfold, and the value of $n$ can only be
$0,\pm 4,...$. In the general case, when the polarizabilities
between the microparticles and ambient fluid is low, the higher
magnetic moments can be neglected since they contribute less than
5 percent of the total energy.$^{12}$ In this picture, the
nonmagnetic microparticle pair reflects the dipole-dipole and
dipole-multipole nature of the interaction, and can be used to
predict the behavior of microparticle chains in simple crystals.

\section{Possible ground state for uniform ordered configurations}\label{secIII}

Let us first consider a bidisperse model which has been widely
used in the study of magnetorheological fluids and ferrofluids.
The model has large amount of spherical nonmagnetic microparticles
with two different sizes suspended in a ferrofluid which is
confined between two infinite parallel nonmagnetic plates with
positions at $z=0$ and $z=L$, respectively. When a magnetic field
is applied, dipole and multipole moments will be induced to appear
in the spheres. The inverse ferrofluid systems consist of
spherical nonmagnetic microparticles in a carrier ferrofluid, and
the viscosity of the whole inverse ferrofluid increases
dramatically in the presence of an applied magnetic field. If the
magnetic field exceeds a critical value, the system turns into a
solid whose yield stress increases as the exerting field is
further strengthened. The induced solid structure is supposed to
be the configuration minimizing the interaction energy, and here
we assume first that the microparticles with two different size
have a fixed distribution as discussed below.

Using the cylindrical coordinates, the interaction energy between
two microparticles labelled as $i$ and $j$ considering both the
dipole-dipole and dipole-multipole effects can be written as
\begin{eqnarray}
U_{ij}(\rho,z)
=\mu(1+\sum_l\frac{f(l)}{r_{ij}^{2l+1}})\cdot(\frac{1-3\cos^2\theta}{r_{ij}^3}),
\end{eqnarray}
where the center-to-center separation
$r_{ij}=|r_i-r_j|=[\rho^2+(z_i-z_j)^2]^\frac{1}{2}$,
 and $\theta$ is
the angle between the field and separation vector $r_{ij}$ (see
Figure 1). Here $\rho=[(x_i-x_j)^2+(y_i-y_j)^2]^\frac{1}{2}$
stands for the distance between chain $A$ and chain $B$ (Figure
2), and $z_i$ denotes the vertical shift of the position of
microparticles. Since the inverse ferrofluid is confined between
two plates, the microparticle dipole at $(x,y,z)$ and its images
at $(x,y,2Lj\pm z)$ for $j=\pm1,\pm2,...$ constitute an infinite
chain. In this work, we would discuss the physical infinite
chains. After applying a strong magnetic field, the mismatch
between the spheres and the host ferrofluid, as well as the
different sizes of the two sorts of spheres will make the spheres
aggregate into lattices like a bct (body-centered tetragonal)
lattice. In fact, the bct lattice can be regarded as a compound of
chains of A and B, where chains B are obtained from chains A by
shifting a distance $r_1$ (microparticle radius) in the field
direction. Thus, we shall study the case in which the identical
nonmagnetic microparticles gather together to form a uniform
chain, when phase separation or transition happens. For long range
interactions, the individual colloidal microparticles can be made
nontouching when they are charged and stabilized by electric or
magnetic static forces, with a low volume fraction of nonmagnetic
microparticles. The interaction energy between the nonmagnetic
microparticles can be divided into two parts: one is from the self
energy of one chain($U_{s}$), the other is from the interaction
between different chains($U_{ij}({\rho,z})$). Consider the
nonmagnetic microparticles along one chain at $r_j=2aj\hat{z}$
$(j=0,\pm1,\pm2,...)$ (namely, chain A), and the other chain at
$r_j=(2j+1)a\hat{z}$ (chain B), the average self energy per
microparticle in an infinite chain is
$U_s=-\mu\sum^{\infty}_{s=1}[\frac{1}{(2as)^3}+2\sum_l\frac{f(l)}{(2as)^{2l+4}}].$

If we notice that for an infinite chain all even mulitipole
contributions vanish due to spatial magnetic antisymmetry around
the spheres, the sum starts at $l=3$. Because the radius of the
sphere is smaller than the lattice parameter $a$, for large
multipole moment, $\frac{r_1^{2l+1}+r_2^{2l+1}}{(2a)^{2l}}\ll1$,
we need only consider the first two moment contributions for
simplicity. Thus the average self energy $U_s$ can be calculated
as
$U_s=-2\mu(\frac{\zeta(3)}{{(2a)}^3}+\frac{f(3)\zeta(6)}{(2a)^6}+\frac{f(5)\zeta(10)}{(2a)^{10}})
=-\mu(\frac{0.300514}{a^3}+\frac{0.0317920f(3)}{a^6}+\frac{0.00195507f(5)}{a^{10}})
,$ where $\zeta(n)=\sum^{\infty}_{s=1}\frac{1}{s^n}$ is the
Riemann $\zeta$ function. The interaction energy between two
parallel infinite chains can be given by
$\frac{1}{2}U_{ij}(\rho,z)$, in which the microparticles along one
chain locate at $r_j=2aj\hat{z}$ $(j=0,\pm1,\pm2,...)$ and one
microparticle locates at $r_j=\rho+z\hat{z}$,
\begin{eqnarray}
U_{ij}(\rho,z)
&=&-\mu[(2+\rho\frac{\partial}{\partial\rho})\sum^{\infty}_{j=-\infty}\frac
{1}{[\rho^2+(z-2ja)^2]^\frac{3}{2}}]\nonumber\\&&-\mu[\sum_lf(l)(2+\frac{3}{2l+2}\rho\frac{\partial}{\partial\rho})
\sum^{\infty}_{j=-\infty}\frac
{1}{[\rho^2+(z-2ja)^2]^{l+2}}]\nonumber\\
&=&U_1+U_2 .
\end{eqnarray}

Following the Fourier expanding technique which is proposed by Tao
{\it et al.},$^{27}$ we derive $U_2$ which is the second part of
$U_{ij}(\rho,z)$ as
\begin{equation}
U_2=-\mu\sum_l\frac{f(l)}{4a\rho^{2l+3}}[-\frac{(2l+1)\sqrt{\pi}\Gamma(l+\frac{3}{2})}
{\Gamma(l+3)}+2^{\frac{1}{2}-l}(\frac{s\rho}{a})^{l+\frac{3}{2}}\pi^{l+2}\cos(\frac{s\pi
z}{a})\cdot S]\label{eq6}
\end{equation}
with
\begin{equation}
S=\sum_{s=1}^{\infty}(\frac{K_{\frac{5}{2}}(\frac{s\pi}{\rho})}{\Gamma(l+3)}+
\frac{4K_{\frac{3}{2}}(\frac{s\pi}{\rho})}{\Gamma(l+2)}) .
\end{equation}
Here $K_i(x)$ represents the $i$th order modified Bessel function,
$\Gamma(x)$  the $\Gamma$ function, and $s$ denotes the index in
Fourier transformation.$^{26}$  And the dipole-dipole energy $U_1$
is written as
\begin{equation}
\ U_1=-\frac{\mu}{a^3}\sum_{s=1}^{\infty}2\pi^2s^2K_0
(\frac{s\pi\rho}{a})\cos(\frac{s\pi z}{a}).
\end{equation}
We obtain the expression for $U_{ij}(\rho,z)$, and the interaction
energy per nonmagnetic microparticle $U(\rho,z)$ is
$U_s+\frac{1}{2}\sum_kU_{ij}(\rho,z)$, where $\sum_k$ denotes the
summation over all chains except the considered microparticle. For
the same reason of approximation discussed above, we need only
choose the first two terms $(l=3$ and $l=5)$ in the calculation.

The interaction between chain A and chain B depends on the shift
$z$, the lattice structure and the nonmagnetic microparticle size.
An estimation of the interaction energy per nonmagnetic
microparticle includes the nearest and next-nearest neighboring
chains, here we could discuss three most common lattice structures:
$bct$, $fcc$, and $hcp$ lattices. For the above lattices, their
corresponding energy of $U_{ij}(\rho,z)$ can be respectively
approximated as
$U_{ij,bct}(\rho,z)=4U_{ij}(\sqrt{3}a,z=0)-4U_{ij}(\sqrt{6}a,z=0)$,
$U_{ij,fcc}(\rho,z)=4U_{ij}(\sqrt{3}a,z=0)-2U_{ij}(2a,z=0)$, and
$U_{ij,hcp}(\rho,z)=3U_{ij}(\sqrt{3}a,z=0)-4U_{ij}(2a,z=0)$.

Figure~\ref{fig3} shows, for different lattices, the dependence of
$U_{ij}(\rho,z)$ on the vertical position shift $z$, which
determines whether the interaction is attractive or repulsive.
$U_{ij}(\rho,z)$ reflects the energy difference  between chain A and
chain B for (a) $bct$, (b) $fcc$, and (c) $hcp$ lattices. It is
evident that, for the same lattice structure, $U_{ij}(\rho,z)$ is
minimized when the size difference between chain A and chain B is
the smallest. For the sake of comparison, we also plot the results
obtained by considering the dipole-dipole interaction only.
Comparing the different lattices, we find that the $bct$ lattice
possesses the smallest energy at the equilibrium point, thus being
the most stable.

Figure~\ref{fig4} displays the interaction energy $U(\rho,z)$
 as a function of the lattice constant $a$
for the $bct$ lattice. It is shown that as the lattice constant
increases, the dipole-multipole effect becomes weaker and weaker,
and eventually it reduces to the dipole-dipole effect. In other
words, as the lattice constant  is smaller, one should take into
account the dipole-multipole effect. In this case, the effect of
polydispersity in size can also play an important role.

Figure~\ref{fig5} displays the interaction energy per nonmagnetic
microparticle $U(\rho,z)$ vs the lattice parameter $a$ for
different lattice structures. The $bct$ structure also proves to
be the most stable state while the $hcp$ lattice has the highest
energy. It also shows that the energy gap between $bct$ lattice
and $fcc$ lattice exists but is small. Figure~\ref{fig6}(a) shows
that the energy gap $\Delta U=U_{bct}-U_{fcc}$ is about 0.5
percent of the interaction energy value. In this aspect, the $bct$
lattice proves always to be a more stable structure comparing with
$fcc$. As the radius of microparticles increases, the energy gap
between $bct$ and $fcc$ lattice enlarges accordingly. That is, the
$bct$ lattice becomes much more stable. Figure~\ref{fig6}(b) shows
the $bct$ lattice energy $U(\rho,z)$ in respect of different sizes
of microparticles for chain A and chain B. It can be seen that the
close touching packing $(r_1=r_2=a)$ has the lowest energy state.
However, also from the graph, the crystal with the same
microparticle size (monodisperse system) may not be the lowest
energy state, which gives a possible way of fabricating different
crystals by tuning the distribution of microparticle size.

\section{Polydisperse system with random
distributions}\label{secIV}

In Section~\ref{secIII}, we have discussed the structure and
interaction in a bidisperse inverse ferrofluid (namely, containing
microparticles with two different sizes). But the interaction form
in polydisperse crystal system is complex and sensitive to the
microstructure in the process of crystal formation. Now we
investigate the structure of polydisperse inverse ferrofluids with
microparticles of different sizes in a random configuration. To
proceed, we assume that the average  radius $r$ satisfies the
Gaussian distribution
\begin{equation}
P(r)=\frac{1}{\sqrt{2\pi}\sigma}\exp\left(-\frac{(r-r_0)^2}{2\sigma^2}\right),
\end{equation}
where $\sigma$ denotes the standard deviation of the distribution
of microparticle radius, which describes the degree of
polydispersity. Integrating eq~\ref{eq6} by $r_1$ and $r_2$, we
could get the average dipole-multipole energy $\overline{U}_2$.
Doing the same calculation to self energy $U_s$, we can get the
average interaction energy
$\overline{U}(\rho,z)=\overline{U}_s+U_1+\overline{U}_2$, where
the microparticle size $r_1$ and $r_2$ are replaced by the mean
radius $r_0$. The microparticle sizes will be distributed in a
wider range as long as a larger $\sigma$ is chosen.

Figure~\ref{fig7}(a) shows the ground state interaction energy of
$bct$ lattice for different polydisperse distributions. As the
degree of polydispersity $\sigma$ increases, the energy $U(\rho,z)$
drops fast, especially when the distribution of microparticle size
gathers around $r=a$. It shows that the inverse ferrofluid crystal
in the formation of ground state tends to include microparticles
possessing more different sizes. The crystal configuration energy of
two uniform chains with identical microparticle aggregation is also
plotted in the graph. Here we consider two cases. First, we assume
the microparticles in chain A and chain B are identical, $r_1 = r_2
= r_0$. As $r_0$ increases, the behavior of energy decreasing is
discovered to be similar with the random configuration with
$\sigma=0.2 r_0$. Second, we set for one chain, such as chain B, the
microparticle size $r_2 = a$ to be unchanged, while the
microparticle size $r_1$ of chain A increases. It shows that the
$bct$ lattice energy for the second case is lower than the first
case and  two other random configurations. And it also shows that
the random configuration is not always the state with the lowest
energy. It proves that polydisperse systems are sensitive to many
factors which can determine the microstructure. Figure~\ref{fig7}(b)
shows the energy gap between bct and fcc lattices for different
distribution deviation $\sigma$. It is evident that higher $\sigma$
leads to larger energy difference between bct and fcc, especially at
larger $r_0$. In other words, at larger $r_0$ and/or $\sigma$, $bct$
lattices are much more stable than $fcc$.

\section{Molecular-dynamic simulations}\label{secV}

Here we use a molecular-dynamic simulation, which was proposed by
Tao {\it et al.},$^{28}$ in order  to briefly discuss the
structure formation of bidisperse inverse ferrofluids. The
simulation herein involves  dipolar forces, multipole forces,
viscous drag forces and the Brownian force. The microparticles are
confined in a cell between two parallel magnetic pole plates, and
they are randomly distributed initially, as shown in Figure 8(a).
The motion of a microparticle $i$ is described by  a Langevin
equation,
\begin{equation}
m_a\frac{d^2\vec{r}_i}{dt^2}=\vec{F}_i-3\pi\sigma\eta\frac{d\vec{r}_i}{dt}+\vec{R}_i(t),
\end{equation}
where the second term in the right-hand side is the Stokes's drag
force, $R_i$ is the Brownian force, and
\begin{equation}
F_i=\sum_{i\neq j}(f_{ij}+f_{ij}^{rep})+f_i^{wall}.\label{sim}
\end{equation}
Here $f_{ij}=-\nabla U(\rho,z)$, while $f_{ij}^{rep}$,
$f_{i}^{wall}$ and $R_i(t)$, have the similar expressions as those
in ref 27 and the references therein. In eq 13, $m_a$ and $\sigma$
are respectively the average mass and diameter of microparticles.
Figure~\ref{fig8} shows the inverse ferrofluid structure with the
parameters, magnetic field $H = 14\,$Oe, temperature $T = 300\,$K,
$A' \equiv \frac{\mu_fm_1m_2}{3\pi^2\sigma^6\eta^2} m_a = 10^{-2},$
and $B'\equiv\frac{\sqrt{6\pi k_B T\sigma^9\eta/\tau}}{3\mu_fm_1m_2}
= 10^{-4}$. Here $A^{'}$ denotes the ratio of a dipolar force to a
viscous force,  $B^{'}$ the ratio of Brownian force to a dipolar
force,  $k_B$ the Boltzmann constant, and $\tau$  the subinterval
time step.

We  take into account a bidisperses system that contains two kinds
of microparticles with different  sizes, as shown in Figure 8. In
details, this figure displays the configuration of microparticle
distribution in a bidisperse system at (a) the initial state, (b)
the state after 15 000 time steps, and (c) the state after 80 000
time steps. The order of Figure 8(c) is better than Figure 8(b).
Here we should remark that 80 000 steps give the sufficient long
time steps to reach the equilibrium state for the case of our
interest. The structure for the bi-disperse system in Figure 8(b)
and (c) has the following features: (i) In the field direction, the
large spheres form the main chains from one plate to the other,
where the large spheres touch each other. (ii) The large spheres
also form many small $bct$ lattice grains. However, they do not form
a large $bct$ lattice. (iii) The small spheres fill the gaps between
these $bct$ lattice grains. From Figure 8, it is observed that, for
the parameters currently used, the order of a bidisperse system
(which is a $bct$-like structure) is not as good as that of
monodisperse system (no configurations shown herein). We should
remark that the long-range interaction can yield the above-mentioned
bct lattice structure, but some perturbations caused by the Brownian
movement existing in the system can change it to another lattice
structure which has similar free energy. Therefore, for the
bi-disperse system of our interest, the large spheres form the main
chains from one plate to the other in the field direction, thus
forming many small bct-like lattice structures. While small spheres
fill the gaps between these bct-like lattices, they themselves do
not form a bct-like lattice due to such perturbations. Here we
should also mention that the degree of order of a specific system
depends on the choice of various physical parameters, for example,
the size of microparticles and so forth.

\section{Summary}\label{secVI}

In summary, by using theoretical analysis and molecular dynamics
simulations, we investigate the structure of colloidal crystals
formed by nonmagnetic microparticles (or magnetic holes) suspended
in a host ferrofluid, by taking into account the effect of
polydispersity in size of the nonmagnetic microparticles. We obtain
an analytical expression for the interaction energy of monodisperse,
bidisperse, and polydisperse inverse ferrofluids. The $bct$ lattices
are shown to possess the lowest energy when compared with other
sorts of lattices, and thus serve as the ground state of the
systems. Also, the effect of microparticle size distributions
(namely, polydispersity in size) plays an important role in the
formation of various kinds of structural configurations. Thus, it
seems possible to fabricate colloidal crystals by choosing
appropriate polydispersity in size. As a matter of fact, it is
straightforward to extend the present model to more ordered periodic
systems,$^{29}$ in which the commensurate spacings can be chosen as
equal or different.

\section*{Acknowledgments}

Two of us (Y.C.J. and J.P.H.) are grateful to Dr. Hua Sun for
valuable discussion. This work was supported by the National
Natural Science Foundation of China under Grant No.~10604014, by
the Shanghai Education Committee and the Shanghai Education
Development Foundation ("Shu Guang" project), by the Pujiang
Talent Project (No. 06PJ14006) of the Shanghai Science and
Technology Committee, and by Chinese National Key Basic Research
Special Fund under Grant No. 2006CB921706. Y.C.J. acknowledges the
financial support by Tang Research Funds of Fudan University, and
by the "Chun Tsung" Scholar Program of Fudan University.

\newpage
\section*{References and Notes}
(1) Odenbach S., {\it Magnetoviscous Effects in Ferrofluids}
Springer, Berlin, 2002.

(2) Meriguet G.; Cousin F.; Dubois E.; Boue F.; Cebers A.; Farago
B.; and Perzynski W. \textit{J. Phys. Chem. B} \textbf{2006}, 110,
4378.

(3) Sahoo Y.; Goodarzi A.; Swihart M. T.; Ohulchanskyy T. Y.; Kaur
N.; Furlani E. P.; and Prasad P. N. \textit{J. Phys. Chem. B}
\textbf{2005}, 109, 3879.

(4) Toussaint R.; Akselvoll J.; Helgesen G.; Skjeltorp A. T.; and
Flekk${\o}$y E. G. \textit{Phys. Rev. E.} \textbf{2004}, 69, 011407.

(5) Skjeltorp  A. T. \textit{Phys. Rev. Lett.} \textbf{1983}, 51,
2306.

(6) Zubarev A. Y.  and Iskakova L. Y. \textit{Physica A}
\textbf{2003}, 335, 314.

(7) Chantrell R. and  Wohlfart E. \textit{J. Magn. Magn. Mater.}
\textbf{1983}, 40, 1.

(8) Rosensweig R. E. \textit{Annu. Rev. Fluid Mech.} \textbf{1987},
19, 437.

(9) Ugelstad J.  et al, \textit{Blood Purif.} \textbf{1993}, 11,
349.

(10) Hayter J. B.; Pynn R.; Charles S.; Skjeltorp A. T.; Trewhella
J.; Stubbs G.; and Timmins P. \textit{Phys. Rev. Lett.}
\textbf{1989}, 62, 1667.

(11) Koenig A.; H\'{e}braud P.; Gosse C.; Dreyfus R.; Baudry J.;
Bertrand E.; and Bibette J. \textit{Phys. Rev. Lett.} \textbf{2005},
95, 128301.

(12) Zhang H. and Widom M. \textit{Phys. Rev. E.} \textbf{1995},
51, 2099.

(13) Friedberg R.  and Yu Y. K. \textit{Phys. Rev. B.}
\textbf{1992}, 46, 6582.

(14) Clercx H. J. H. and Bossis G. \textit{Phys. Rev. B.}
\textbf{1993}, 48,2721.

(15) Mondain-Monval O.; Leal-Calderon F.; Philip J.; and Bibette J.
\textit{Phys. Rev. Lett.} \textbf{1995}, 75, 3364.

(16) Sacanna S.  and Philipse A. P. \textit{Langmuir} \textbf{2006},
22, 10209.

(17) Claesson E. M.  and Philipse A. P. \textit{Langmuir}
\textbf{2005}, 21, 9412.

(18) Dai Q. Q.; Li D. M.; Chen H. Y.; Kan S. H.; Li H. D.; Gao S.
Y.; Hou Y. Y.; Liu B. B. and Zou G. T. \textit{J. Phys. Chem. B}
\textbf{2006},110, 16508.

(19) Ethayaraja M.; Dutta K.; and Bandyopadhyaya R. \textit{J. Phys.
Chem. B} \textbf{2006}, 110, 16471.

(20) Jodin L.; Dupuis  A. C.; Rouviere E.; and Reiss P. \textit{J.
Phys. Chem. B} \textbf{2006}, 110, 7328.

(21) Gao L. and Li Z. Y. \textit{J. Appl. Phys.} \textbf{2002}, 91,
2045.

(22) Wei E. B.; Poon Y. M.; Shin  F. G.; and Gu G. Q. \textit{Phys.
Rev. B} \textbf{2006}, 74, 014107.

(23) Krist$\acute{\rm o}$f T.  and  Szalai I. \textit{Phys. Rev. E}
\textbf{2003}, 68, 041109.

(24) Huang J. P. and Holm C. \textit{Phys. Rev. E} \textbf{2004},
70, 061404.

(25) J. \v{C}ern\'{a}k, G. Helgesen, and A. T. Skjeltorp,
\textit{Phys. Rev. E} \textbf{2004}, 70, 031504.

(26) Jackson J. D.,{\it Classical Electrodynamics}, 3rd edition
Wiley, New York, 1999, Chapter 4.


(27) Tao R.  and  Sun J. M. \textit{Phys. Rev. Lett.}
\textbf{1991}, 67, 398.

(28) Tao R.  and  Jiang Q. \textit{Phys. Rev. Lett.}
\textbf{1994}, 73, 205.



(29) Gross M. and Wei C. \textit{Phys. Rev. E} \textbf{2000}, 61,
2099.

\newpage

\section*{Figure captions}

Figure~1.  Schematic graph showing two nonmagnetic microparticles
(magnetic hole) of radius $r_1$ and $r_2$, suspended in a
ferrofluid under an applied magnetic field $H$.

Figure~2.  Three different lattices, $bct$, $fcc$, and $hcp$,
which are composed of non-touching microparticles with different
size distribution.

Figure~3.  The dependence of interaction energy $U_I(\rho,z)$ (in
units of $\mu_0$) versus vertical shift $z$ for different
lattices: (a) $bct$, (b) $fcc$, and (c) $hcp$. In the legend,
"dipole-dipole" denotes the case that the dipole-dipole
interaction is only considered  for calculating the interaction
energy.

Figure~4.  The interaction energy $U(\rho,z)$ versus lattice
constant $a$. The solid line stands for the case in which the
dipole-dipole interaction is only considered.

Figure~5.  The interaction energy per microparticle $U(\rho,z)$
versus lattice constant $a$ for different lattices, $bct$, $fcc$,
and $hcp$.

Figure~6.  (a) The energy gap $\Delta U = U_{bct}-U_{fcc}$ (in
units of $\mu_0$) versus different size of nonmagnetic
microparticles in chain A and chain B; (b) The $bct$ lattice
energy $U_(\rho,z)$ (in units of $\mu_0$) versus different sizes
of nonmagnetic microparticles in chain A and chain B.

Figure~7.  (a) The ground state interaction energy of a $bct$
lattice versus microparticle size $r_0$ for random polydispersity
configuration (solid, dashed, and dotted lines) and a
configuration composed of two different uniform chains
(dash-dotted and short-dash-dotted lines). (b) The energy gap
$\Delta U$ between $bct$ and $fcc$ lattices for random
polydispersity versus $r_0$.

Figure~8.  The  configuration of nonmagnetic microparticle
distribution at (a) the initial state, (b) the state after 15000
time steps, and (c) the state after 80000 time steps.

\clearpage
\newpage

\begin{figure}[h]
\includegraphics[width=400pt]{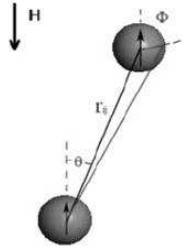}
\caption{Jian, Gao, Huang, and Tao}\label{fig1}
\end{figure}
\clearpage
\newpage

\begin{figure}[h]
\includegraphics[width=400pt]{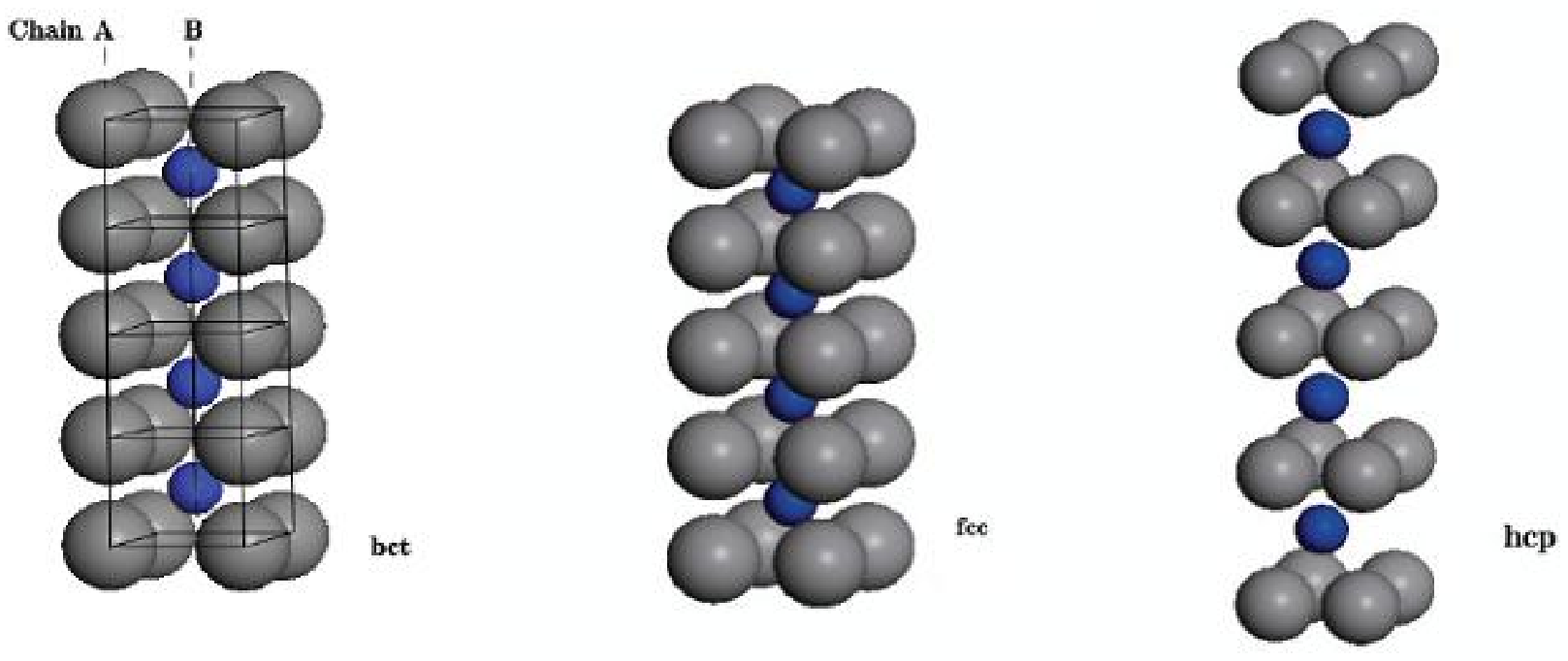}
\caption{Jian, Gao, Huang, and Tao}\label{fig2}
\end{figure}
\clearpage
\newpage

\begin{figure}[h]
\includegraphics[width=400pt]{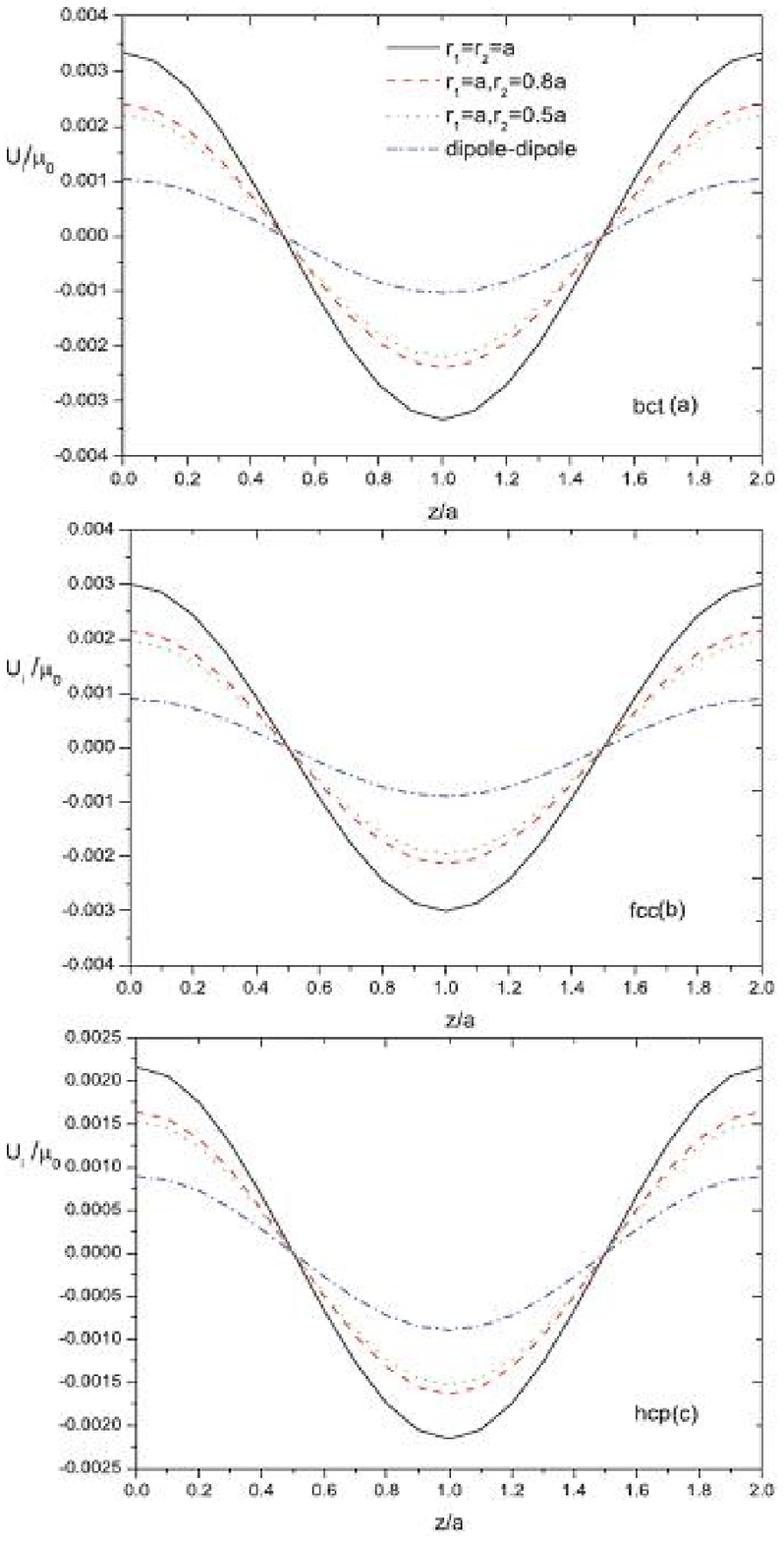}
\caption{Jian, Gao, Huang, and Tao}\label{fig3}
\end{figure}
\clearpage
\newpage

\begin{figure}[h]
\includegraphics[width=400pt]{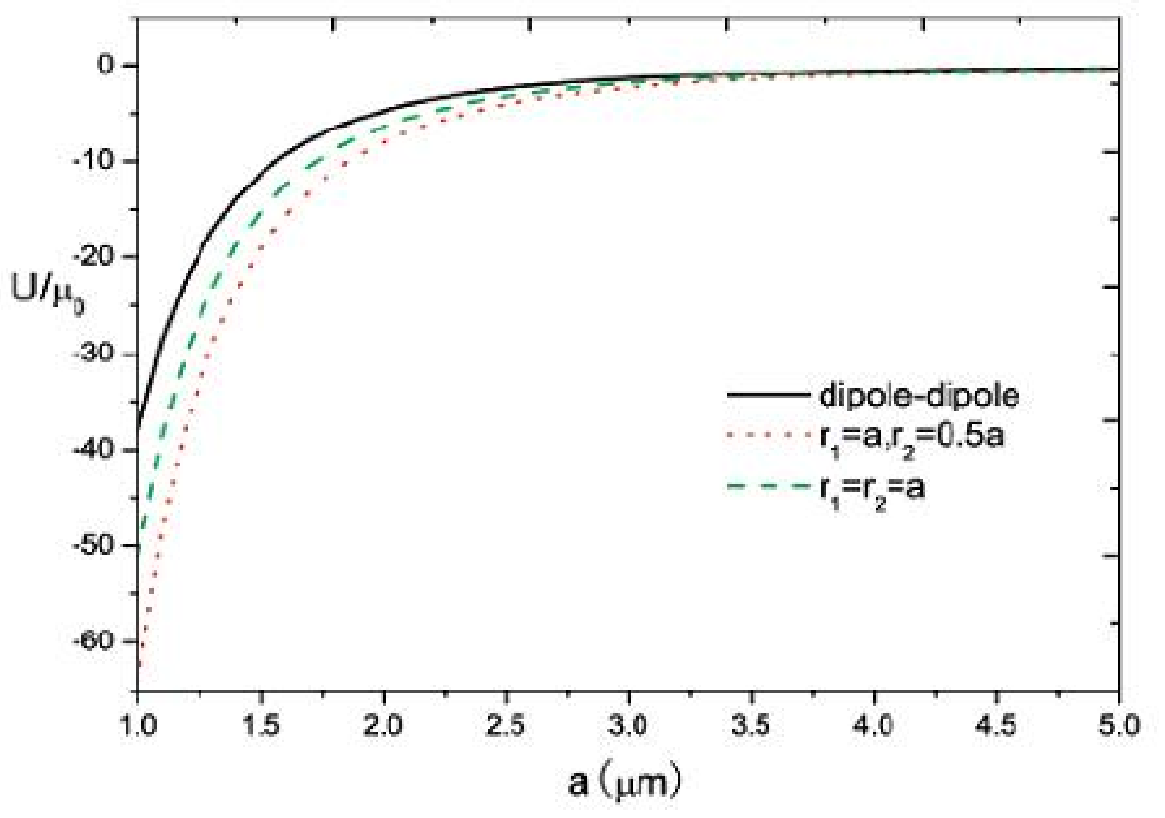}
\caption{Jian, Gao, Huang, and Tao}\label{fig4}
\end{figure}
\clearpage
\newpage

\begin{figure}[h]
\includegraphics[width=400pt]{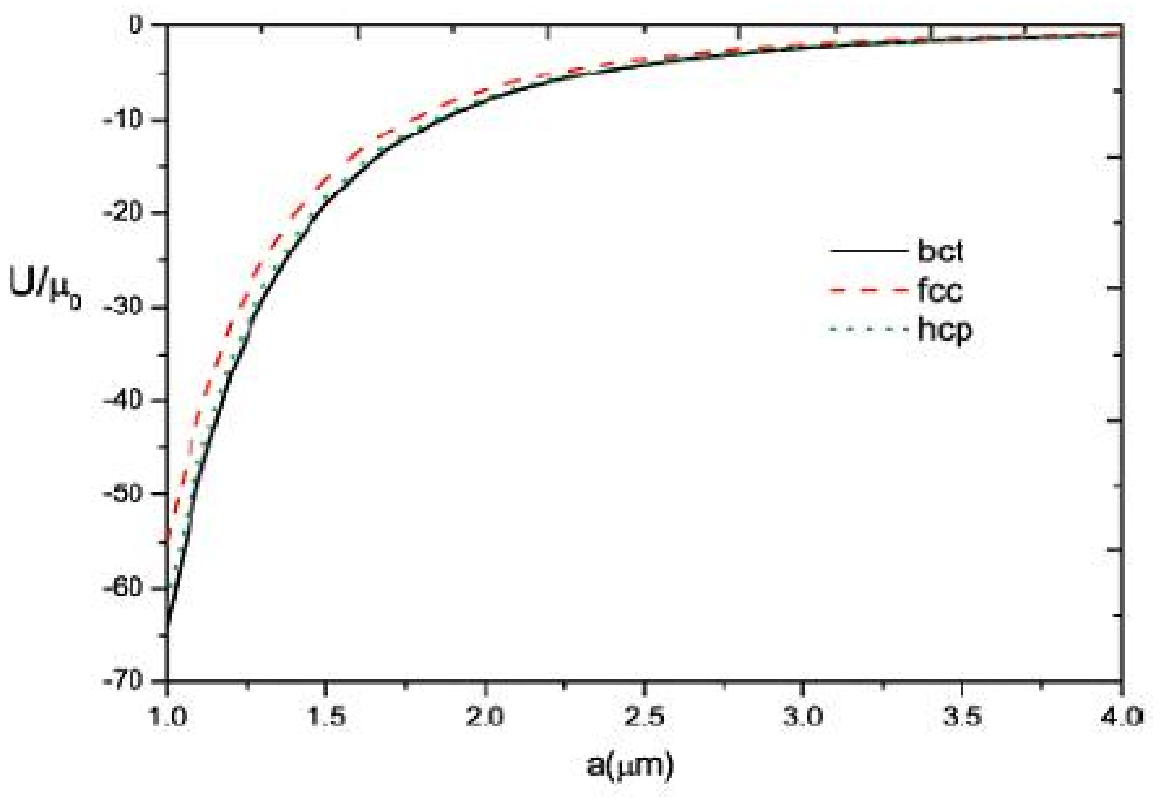}
\caption{Jian, Gao, Huang, and Tao}\label{fig5}
\end{figure}
\clearpage
\newpage

\begin{figure}[h]
\includegraphics[width=400pt]{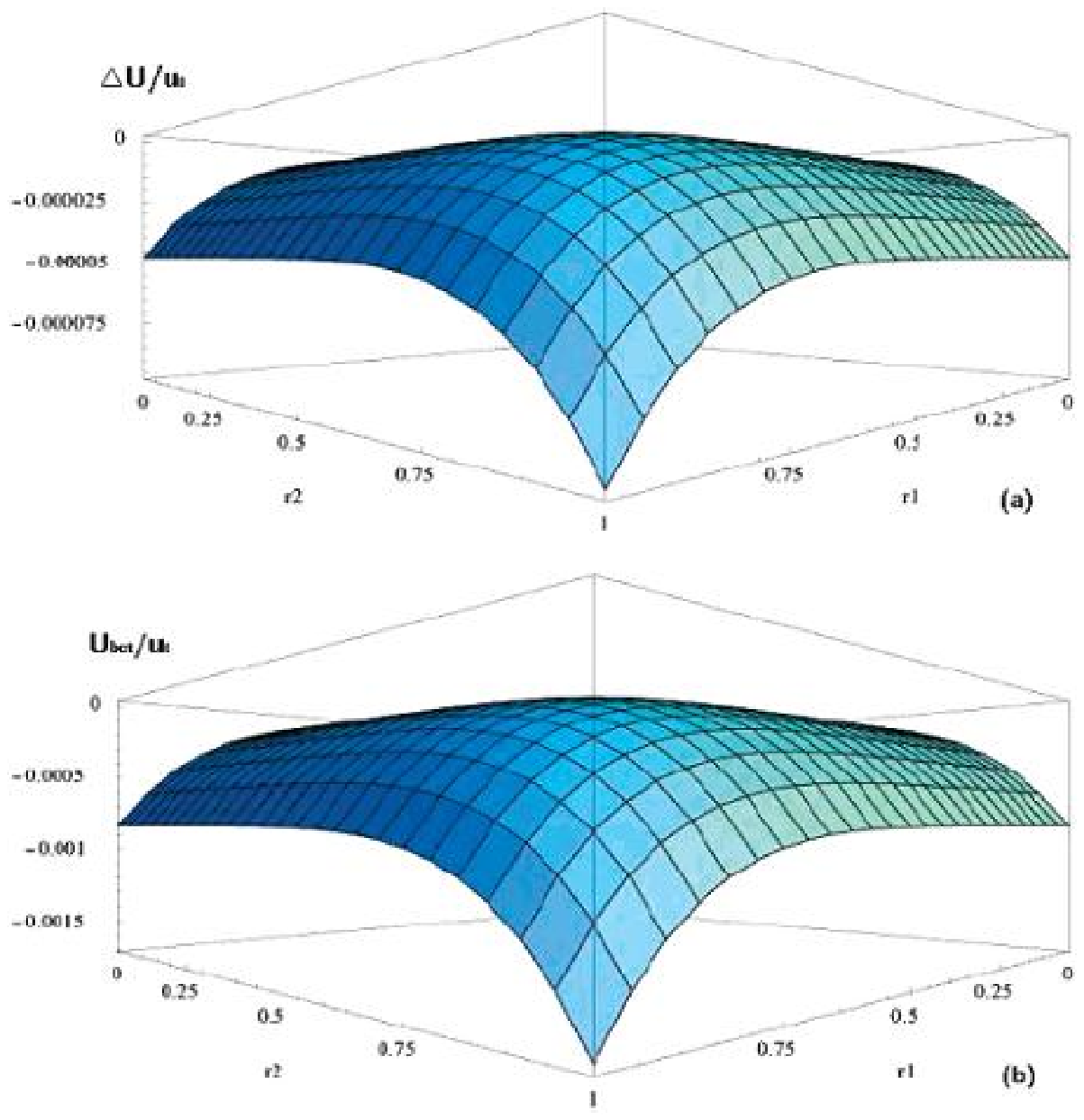}
\caption{Jian, Gao, Huang, and Tao}\label{fig6}
\end{figure}
\clearpage
\newpage

\begin{figure}[h]
\includegraphics[width=400pt]{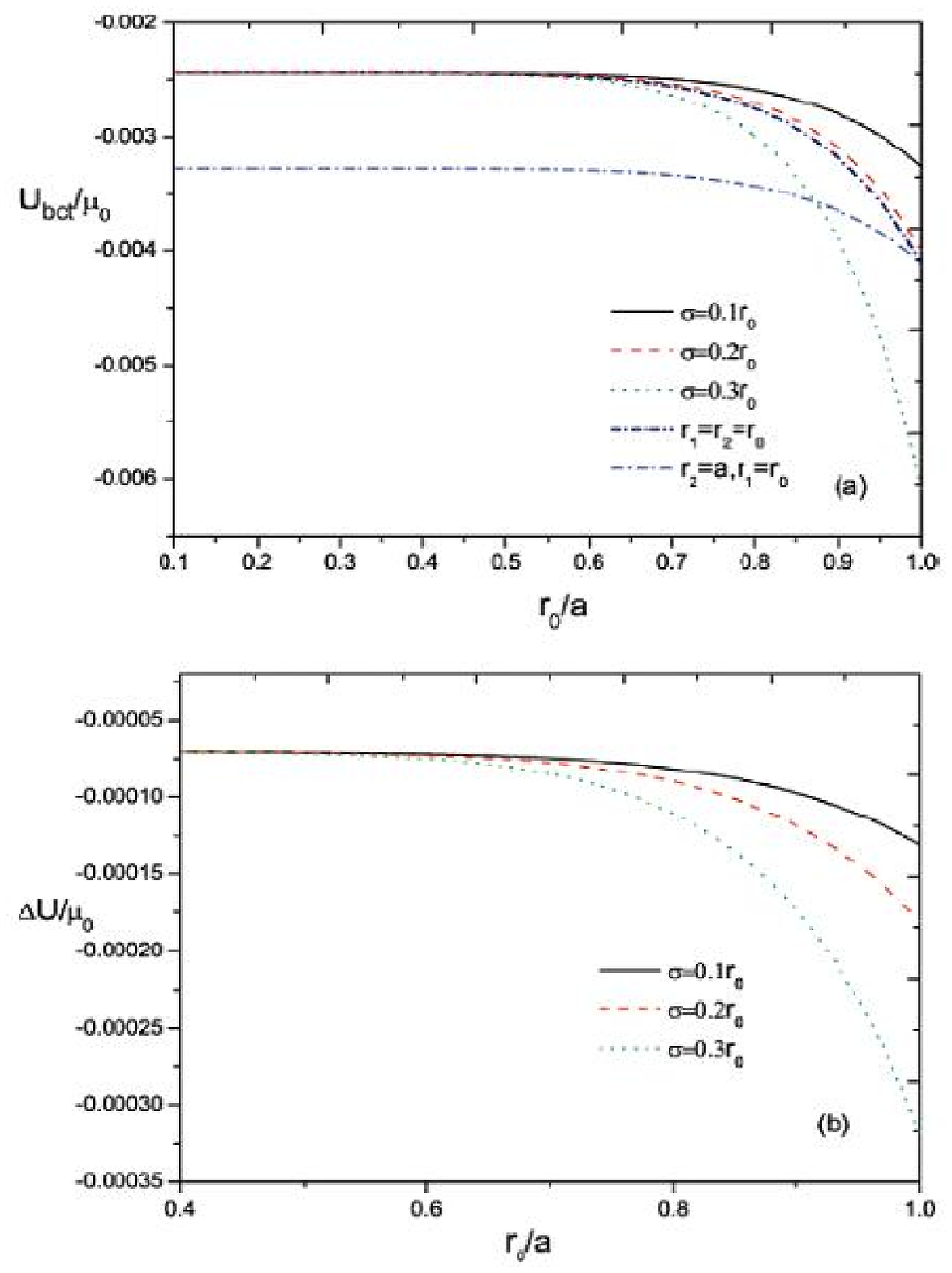}
\caption{Jian, Gao, Huang, and Tao}\label{fig7}
\end{figure}
\clearpage
\newpage

\begin{figure}[h]s
\includegraphics[width=400pt]{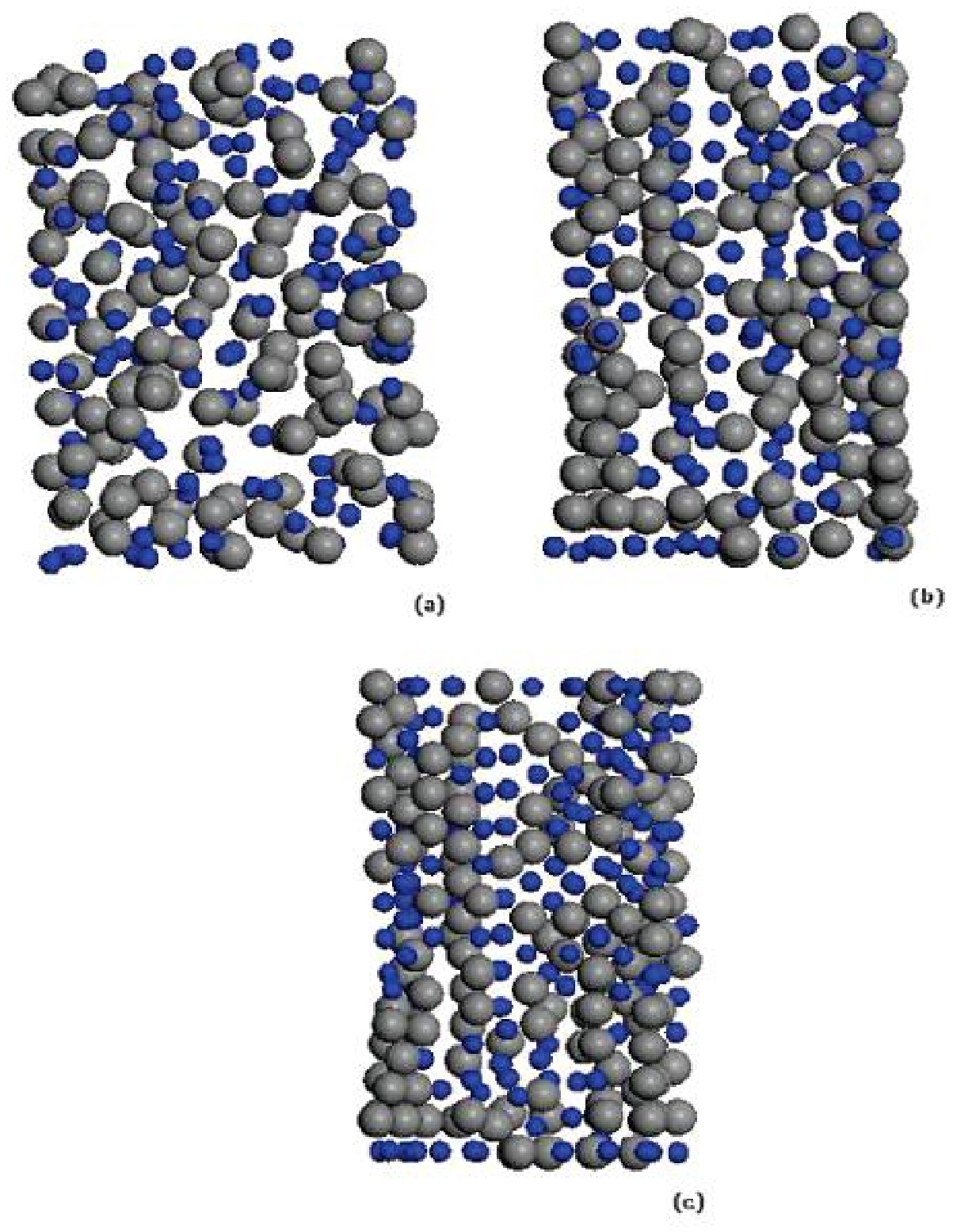}
\caption{Jian, Gao, Huang, and Tao}\label{fig8}
\end{figure}
\end{document}